# Pauli spin blockade at room temperature in double-quantum-dot tunneling through individual deep dopants in silicon

Yoshisuke Ban[1], Kimihiko Kato[2], Shota Iizuka[2], Hiroshi Oka[2], Shigenori Murakami[2], Koji Ishibashi[1,3], Satoshi Moriyama[4], Takahiro Mori[2], Keiji Ono[1,3*]

[1]Advanced Device Laboratory, RIKEN; Wako, Saitama 351-0198, Japan.
[2]Semiconductor Frontier Research Center (SFRC), National Institute of Advanced Industrial Science and Technology (AIST); Tsukuba, Ibaraki 305-8568, Japan.

[3]Center for Emergent Matter Science (CEMS), RIKEN; Wako, Saitama 351-0198, Japan.

[4]Department of Electrical and Electronic Eng

ineering, Tokyo Denki University; Adachi, Tokyo 120-8551, Japan.

* Email: k-ono@riken.jp

## Abstract

Pauli spin blockade (PSB) is a spin-dependent charge transport process that typically appears in double quantum dot (QD) devices and is employed in fundamental research on single spins in nanostructures to read out semiconductor qubits. The operating temperature of PSB is limited by that of the QDs and remains below 10 K, limiting wide application development. Herein, we confirm that a single deep dopant in the channel of a silicon field effect transistor functions as a room-temperature QD; consequently, transport through two different deep dopants exhibits PSB up to room temperature. The characteristic magnetoconductance provides a means to identify PSB and enables the PSB device to function as a magnetic sensor with a sensitivity below geomagnetic field. Lifting in PSB caused by magnetic resonance (50 K) and Rabi oscillations (10 K) are also observed. Further development of this unique system may lead to room-temperature quantum technologies based on silicon technology.

## Introduction

Pauli spin blockade (PSB) is a universal phenomenon observed in the charge transport of double quantum dot (QD) devices[1]. Owing to Pauli's exclusion principle, charge transport is blocked by the alignment of the spins of each dot. This has been observed in devices with various materials/structures, such as GaAs[1–3], carbon nanotubes[4–6], semiconductor nanowires[7–14], and Si[15–24].

The universality of PSB is attributed to the minimum numbers of the components and the physical conditions (mentioned later) they satisfy[1–24]. The building blocks comprise two QDs and source/drain electrodes that are weakly tunnel-coupled in series. The conditions that must be satisfied are as follows: the tunneling causes a charge transport cycle *01-11-02-01…* (the first and second numbers indicate the number of charges for each dot), and the confinement strength (quantization level spacing and on-site Coulomb energy) of the dots is sufficiently larger than the thermal energy. According to Pauli's exclusion principle, the *02*-state can take only a spin singlet state; however, the *11*-state can take either a spin singlet or triplet state. Thus, charge transport is eventually blocked by the spin-triplet *11*-state.

If the spin-triplet states are scattered into (or mixed with) the spin-singlet state, the PSB is lifted, resulting in a finite leakage current. This leakage current facilitates the investigating the physics of the dot spin via charge transport. The fluctuations/polarizations of the nuclear spins and spin–orbit interaction (SOI) in the dots have been investigated using the PSB leakage current[2, 5, 7–11, 13–19, 21–23, 25–29]. In particular, a characteristic peak and/or dip structure appears near a region wherein the magnetic field is zero when assessing the magnetic field dependence of the leakage current. Because the PSB is lifted by rotating one of the dot spins using magnetic resonance, the PSB leakage current under magnetic resonance conditions is used as a time-ensemble readout for the spin qubit[3, 6, 8, 13, 18–20].

Conversely, in systems wherein the charge and/or spin states cannot be sufficiently evaluated, observing the aforementioned characteristics facilitates the identification of PSB. PSB has been adopted as a model for the room temperature low-field magnetoresistance effect observed in $\pi$-conjugated organic materials[30].

Owing to the confinement of QDs (approximately equal to 5 meV in typical cases[1–24]), their operating temperature, as well as the operation temperature of PSB for double dot, are typically of the order of 1 K (one order of magnitude smaller thermal energy than the confinement energy). PSB research has largely focused on the integration of QDs at low temperatures, and no research has focused on further improving the operating temperature, except a study by Ono et al.[20], which is explained in the following.

Deep dopants in semiconductors create deep dopant levels in the band gap of the semiconductor, with specific depth in the band gap and a small depth distribution, achieving stronger confinement (typically of order of 0.3 eV, as described subsequently) than those in QDs and/or shallow dopants (order of 10 meV, for P, As, and B) [20, 31–37]. Deep dopants have been used instead of QDs[20]. An Al-N pair in Si (adjacent Al and N form a single dopant level) was used as a deep dopant. A short-channel tunnel field effect transistor (TFET) structure was adopted to realize the tunneling transport through a deep dopant level. A TFET has an N+ source and P+ drain electrodes and thus forms a planar p-type intrinsic n-type (PIN) structure with a gate electrode. In short-channel TFETs, dopant levels deep in the bandgap can be probed by PIN tunneling current (source–drain current). This type of tunneling current is not possible with ordinary FETs. In this TFET, single-charge tunneling (source–deep dopant–drain) at 1 K up to room temperature (300 K) has been observed. At low temperatures, double-quantum-dot tunneling transport occurs because of the combination of the deep dopant and an unintentionally introduced shallow dopant. PSB was observed up to 10 K, limited by the thermal excitation of the shallow dopant that allow occupation of the *02*-state with a spin triplet[20]. Thus, by replacing the shallow dopant with a deep one, it is expected that PSB will operate at even higher temperatures.

The PSB for the *11-02* charge states is essentially the same as the spin-dependent tunneling of charge from the spin 1/2 donor to the spin 1/2 acceptor. This is because their spin states can accept a singlet or triplet before recombination and only accept a singlet after recombination. Thus, double-quantum-dot tunneling transport through a deep donor and a deep acceptor in series can facilitate the realization of a room-temperature PSB if both the deep donor and deep acceptor have confinement strengths that are sufficiently larger than the room-temperature thermal energy (26 meV).

Group III and V shallow dopants are often referred to as antihydrogen/hydrogen atoms in silicon, whereas Group II and VI dopants can be considered antihelium/helium in silicon. Each of these strongly binds to (up to two) carriers (electrons or holes) even at room temperature. Among these deep dopants, we focused on S and Zn, which have been studied for a long time and are well understood[31–37]. S and Zn in Si are known to be thermodynamically stable substitutional dopants. Both of their on-site Coulomb energies are approximately 0.3 eV[31–34]. Both of them also exhibit quantization level spacing of approximately 0.3 eV for the ground and first excited levels[35–37]. The spin states of their charged states ($S^+$ and $Zn^-$, where the superscript represents the charge state) have been investigated. $S^+$ has spin 1/2 ground states, and $Zn^-$ has a nearly degenerate ground state manifold comprising spin 1/2 and spin–orbit 3/2 states[38, 39]. Supplementary Note 1 presents details on the ground and excited states of the S and Zn dopants.

Here, we demonstrated room-temperature PSB in a three-step tunneling process through deep levels of S and Zn dopants in Si. This unique system provides probes for a single deep dopant, particularly its spin and nearby nuclear spins, as well as a technological basis for room-temperature silicon quantum technologies such as magnetic sensors.

## Results

### PSB in S/Zn-codoped TFET

Fig. 1 illustrates the strategy for room-temperature PSB. By using a short-channel TFET (Fig. 1a), we doped S and Zn around the TFET channel. Although the density profiles of these deep dopants were controlled by ion-implantation and post-annealing conditions, the locations of the individual dopants were random, and a TFET device with appropriate dopant locations should be post-selected from many TFET devices fabricated with the same doping conditions. Room-temperature PSB can occur for a three-step percolation-like tunneling through a single S and a single Zn dopant, as depicted in Fig. 1b. As shown in the schematic energy diagram, along with the N+ source electrode, channel, and P+ drain electrode (Fig. 1c), the deep dopant levels for $S^+$ and $Zn^-$ are nearly aligned within the transport window of the source/drain electrodes. At the $S^+$ ($Zn^-$) level, a charge transition of the electron (hole) number between zero and one occurs, thus satisfying the aforementioned PSB condition. Because all energy-level spacings (approximately 0.3 eV) are more than 10 times larger than the room-temperature thermal energy (26 meV), the PSB can be observed up to room temperature.

The devices were fabricated within the framework of standard silicon technology, including S and/or Zn (co-)doping. Preliminary research has been conducted for (co-)doping S/Zn into TFETs[40], and the same (co-)doping conditions were employed (See Supplementary Note 2). We adopted a structure wherein part of the source/drain electrode overlapped with the gate electrode, thus defining a short effective channel length, $L_{\mathrm{eff}}$ (designed to range from 14 to 24 nm). The measurement setup (source–drain voltage $V_{\mathrm{SD}}$, gate voltage $V_{\mathrm{G}}$, source–drain current $I_{\mathrm{SD}}$ and magnetic field $B$) is shown in Fig. 1a (See Method and Supplementary Note 3).

### Single-charge tunneling

Fig. 2 displays the results of single-charge tunneling. These results were obtained by using devices in which a two-step tunneling (source–deep dopant–drain) dominated the transport. Fig. 2a shows the $V_{\mathrm{SD}}$ and $V_{\mathrm{G}}$ dependence of the differential conductance $\mathrm{d}I_{\mathrm{SD}}/\mathrm{d}V_{\mathrm{SD}}$ measured at a temperature $T$ of 10 K in a device with only S doped. Coulomb diamonds, characteristic structures for single-charge tunneling, were observed. The width of the diamonds, as observed by $V_{\mathrm{SD}}$, was approximately 0.25 and 0.5 V. Thus, the charging energy was sufficiently larger than the thermal energy at room temperature. Fig. 2b shows a similar plot at room temperature, where Coulomb diamonds were still visible. Figs. 2d and 2e depict similar measurements performed on devices doped only with Zn. Clear Coulomb diamonds of similar sizes at 10 K and room temperature were observed. The transport models are schematically depicted in Figs. 2c and 2f.

As shown in Fig. 2c (2f), each S (Zn) dopant had two levels corresponding to electron (hole) numbers 1 and 2, respectively. Single-charge tunneling through these electron (hole) = 2 levels was observed as a second neighboring Coulomb diamond, the size of which reflected the energy difference between the dopant levels for different charge states. In Fig. 2a, the width of the Coulomb diamond observed at $V_{\mathrm{G}}$ = 0.5–1 V is 0.25 V; this can be interpreted as electron number = 2 diamond. Although the clarity is lower in Fig. 2d than in Fig. 2a, we can interpret hole number = 2 diamond with a width of approximately 0.3 V at $V_{\mathrm{G}}$ = -0.5–0 V (further details in Supplementary Note 4). We evaluated several devices with similar $L_{\mathrm{eff}}$ values, but without deep dopants. However, Coulomb diamonds (similar to those shown in Fig. 2) have never been observed. These facts indicate that the observed single-charge tunneling in Figs. 2a and 2d is owing to the single dopants S and Zn, respectively. These facts also form the basis for room-temperature PSB, which we will discuss subsequently.

Single-charge tunneling at room temperature is achieved by various nanostructures[41, 42]. However, because these nanostructures are defined by their surfaces/interface, the smaller the QD, the closer its surfaces/interface. Single-charge tunneling often becomes unstable owing to the influence of trap sites at the surface/interface. Therefore, even if clean single-dot-like behavior is observed at room temperature, these traps behave as unintended QDs at low temperatures, resulting in multiple-quantum-dot characteristics. Conversely, our method shows clean single-dot characteristics at room and low temperatures. This is because the nanostructures are provided by deep dopants in Si and there are no surface/interfaces in their vicinity. This advantage is also utilized in the formation of double quantum dots, which will be described next.

### Zero-*B* dip structure at 10–300 K

Fig. 3a shows the $\mathrm{d}I_{\mathrm{SD}}/\mathrm{d}V_{\mathrm{SD}}$ intensity plot at 10 K for an S/Zn-codoped TFET, where (as discussed later), three-step tunneling (source–deep dopant 1–deep dopant 2–drain) instead of two-step tunneling dominated the transport. Clear Coulomb diamonds similar to those observed in Fig. 2 were not observed. Instead, the current threshold at $0.2 < V_{SD} < 0.8$ V exhibited a zigzag structure with respect to $V_G$. Negative $dI_{SD}/dV_{SD}$ values were occasionally observed. These characteristics indicate single-charge tunneling through two (or more) QDs in series. Based on these characteristics, it is difficult to identify the charge/spin states of the dopants involved in the tunneling path. However, if the tunneling path exhibits a PSB, we can expect a characteristic $B$ dependence of the PSB leakage current, $I_{SD}$ peak, and/or dip structure near zero $B$[2, 5, 7, 9–11, 13–15, 17, 21, 22, 27–29].

We measured the $B$ dependence of $I_{SD}$ for various sets of ($V_{SD}$, $V_G$) and observed a zero-$B$ dip structure, as shown in Fig. 3b. This dip structure was maintained up to room temperature under certain ($V_{SD}$, $V_G$) conditions that kept the transport conditions the same (Figs. 3c–f; further details in Supplementary Note 5). Because of the small signal ratio and large noise at room temperature, the $B$ dependence of $I_{SD}$ was measured repeatedly for 1000 times and the average was considered. These zero-$B$ dip structures can be fitted with Lorentzian curves predicted by PSB theory under strong SOI conditions[28], suggesting a PSB from low temperatures up to room temperature. The room-temperature PSB in this device can be understood using the scheme shown in Figs. 1b and 1c.

For S/Zn-codoped TFETs with longer $L_{eff}$, the zigzag structure of the current threshold became modest, and the characteristics approached those of ordinary long-channel TFETs. In these long-channel devices, a zero-$B$ dip structure at room temperature has never been observed. This result supports PSB in the three-step percolation-like tunneling model.

Supplementary Note 6 also presents an S/Zn-codoped TFET in which two-step tunneling dominates.

## Zero-*B* dip and/or peak structures at 300 K

A characteristic zero-$B$ dip (and peak) structure at room temperature was also observed in several other S/Zn-codoped TFETs. Fig. 4a shows the zero-$B$ dip structure observed for one of the S/Zn-codoped TFETs with a small $I_{SD}$ of approximately 10 pA. In addition to the zero-$B$ dip at $|B| < 2$ mT, a peak envelope with a width of 22 mT was observed. Detailed measurements at approximately 0.6 mT, at which a steep change of current was observed, are shown in the inset. A magnetic field sensitivity of 20 μT was obtained, which was limited by a current noise of 1 fA. Fig. 4b presents the results for another S/Zn-codoped TFET, which also exhibited dip and peak envelopes. The inset shows the detailed measurement near zero $B$ and reveals the smallest dip width (0.26 mT). The results for five other S/Zn-codoped TFETs are shown in Supplementary Note 7*)*.

We evaluated several dozen S/Zn-codoped TFETs with a similar range of $L_{eff}$ and found that eight devices exhibited room-temperature zero-$B$ structures. Therefore, the device yield for observing zero-$B$ structures was approximately 10 %, which is reasonable considering the random nature of the deep-dopant locations.

We compared the widths of the zero-$B$ dips and peak envelopes observed at room temperature with PSB theory under SOI and nuclear spin fluctuation conditions. The zero-$B$ dip can be explained by coherent spin-flip tunneling induced by SOI[28]. Furthermore, the zero-$B$ peak of PSB leakage current can be explained by random fluctuations of effective nuclear magnetic fields[27]. In addition, coexistence of both was also discussed[29]. A dip/peak structure without nuclear spin fluctuations has also been discussed[10], where a relaxation to the lowest spin triplet state is the origin of the peak envelope. This spin relaxation can be dominant at milli-Kelvin and Tesla regions. However, this condition does not apply in our case (300 K and mT).

As shown in Fig. 4c, an approximately linear correlation is observed over a broad range between the dip widths observed for eight S/Zn-codoped TFETs and their $I_{SD}$ values. This tendency was consistent with the theory[28, 29] that the dip width is proportional to the tunneling rate between the two QDs (dopants) and $I_{SD}$. The deviation in the plot for the smallest width toward the high $I_{SD}$ side can be attributed to a small (pA-level) gate leakage current and/or a weak source–drain shunt current superposed onto the $I_{SD}$.

We estimated the nuclear fluctuations in the S dopant based on a study of nuclear fluctuations in a typical Si QD[43]. Considering the difference in the confinement strength for a QD (5 meV for in-plane directions and 100 meV in the case of a perpendicular direction to the plane) and the S dopant (300 meV for all directions), we estimated the ratio of their effective volumes to be approximately 100. This ratio resulted in a nuclear fluctuation of approximately 26 mT for the S dopant, which agrees with the observed peak envelope widths of 15–30 mT in Fig. 4c (further details in Supplementary Note 8). Nuclear fluctuations for Zn are expected to be similar to, or smaller than, those for S, reflecting the *p*-orbital nature of the valence band.

**Magnetic resonance and Rabi oscillations**

Magnetic resonance responses and Rabi oscillations were observed at lower temperatures. The results are summarized in Fig. 5.

Fig. 5a shows the magnetic resonance characteristics of a S/Zn-codoped TFET observed at 30 K. The $V_{SD}$ and $V_G$ were set to the conditions where the zero-$B$ structure was observed (details can be found in Supplementary Note 9), $B$ and the frequency of the alternating current magnetic field were swept, and the change in $I_{SD}$ was intensity plotted. Current peaks due to magnetic resonance were visible at frequencies proportional to $B$ and temperatures up to 50 K (Figs. 5b and 5c). To the best of our knowledge, this is five times the highest temperature ever reported for a magnetic resonance response detected by PSB[20]. The obtained $g$-factor was approximately 2.0, which is consistent with the magnetic resonance experiment performed on the $S^+$ ensemble in Si[38].

Fig. 5d shows the pulsed magnetic resonance observed with another S/Zn device wherein microwaves were pulse-modulated, and the change in $I_{SD}$ is plotted against the pulse length ($l$) and $B$. Rabi oscillations were visible at a temperature of 10 K. To the best of our knowledge, this temperature of 10 K is the same as the highest temperature reported so far for Rabi oscillations detected by PSB, but the oscillations appear more clearly[20]. Fig. 5e shows the Rabi oscillations under a resonant magnetic field at three different microwave powers. Further details can be found in Supplementary Note 9.

These results provide clear evidence for PSB. However, these PSBs, observed only at low temperatures, are thought to be caused not by the combination of deep dopants alone, but by a combination of deep and shallow dopants[20]. At higher temperatures (>50 K), the occupation of the *02*-state with a spin triplet is allowed for shallow dopants, and the low temperature PSB is lifted. We emphasize that the above room temperature PSB is due to the combination of deep dopants, S and Zn, and not due to shallow dopants.

## Discussion

To the best of our knowledge, PSB is the only mechanism that can explain the room temperature zero-$B$ structures. Although the weak localization effect is known to show zero-$B$ dip (or peak) for metallic semiconductors at low temperature, it cannot explain our observations. Weak localization is a quantum interference effect from electron (or hole) orbits in "localization loops" formed by multiple scattering by impurities. If the zero-$B$ dip with a width of 1.5 mT (observed at room temperature) is due to weak localization, the diameter of the localized loop would be as large as 3 μm. It is inconceivable that a loop of this size would exhibit quantum interference effects at room temperature, and there is no room in the device channel to accommodate such a loop.

Although the zero-$B$ structures were observed up to room temperature, the magnetic resonance response was not. Magnetic resonance response was not observed in the device shown in Fig. 3 at 10–300 K, or in the devices shown in Fig. 4 at 300 K. The reasons for the missing magnetic resonance response, strong SOI-induced spin-flip tunnels from the triplet *11*-state to the 02 state, and/or PSB including spin–orbit 3/2 states of $Zn^-$ are discussed in Supplementary Note 10. Magnetic resonance at room temperature is expected to pave the way for room-temperature qubits, such as those used in sensors with diamond nitrogen-vacancy centers.

The magnetic field sensitivity of 20 μT (inset of Fig. 4a) indicates that this device can be potentially used as a sensor capable of detecting geomagnetism (approximately 50 μT). As it is a sensor based on strongly localized spins, it has a potential for high-spatial resolution limited by the size of the deep dopant. Because the device yield is limited, as mentioned above, this device is more suitable for individual applications, such as sensors, rather than for integrated systems. Moreover, as it can be fabricated using existing silicon technology, it holds potential for industrial applications. We reserve the analysis of the noise spectral-density (in units of μT/√Hz) based on time-series data for the future (see Supplementary Note 11).

Organic semiconductors can be viewed as random networks consisting of a macroscopic number of QDs, and the PSBs in the tunneling between various QD pairs are assumed to be the origin of the room temperature magnetoresistance effect[30]. By contrast, it is to be noted that our room-temperature PSB devices exhibit percolation conduction based on the use of only two QDs (dopants). The former deals with the statistical average of PSBs in a collective system, whereas the latter deals with PSBs in a single dopant pair. This was achieved by the established manufacturing process for silicon at the nanometer scale, established physics of single-charge tunneling, and evaluation methods based on it. The non-ensemble nature of our system is manifested in the $I_{SD}$ dependence of the zero-$B$ dip width, as discussed in Fig. 4c. The $I_{SD}$ corresponds to the tunneling rate, i.e., the distance between dopant pairs, and changes in zero-$B$ structure depending on this distance is illustrated. These non-ensemble results enable discussion of the effects of SOI and nuclear spin fluctuations separately. These results cannot be observed in an ensemble system with various QD-pair distances. Furthermore, theory[28, 29] predicts a "magic direction" for the magnetic field at which the zero-$B$ dip structure disappears. This direction is related to an effective SOI magnetic field for a particular dopant pair. It would be difficult to observe this magic direction in the ensemble system where the effective SOI magnetic field is randomly distributed for various QD pairs. To the best of our knowledge, no magnetic field direction has been reported in which the zero-$B$ structure disappears in the magnetoresistance effect of organic semiconductors. Observing the magic direction in our system would be interesting future work.

An obvious limitation of this system is that the location of deep dopant cannot be individually controlled. Atomic-level positional control of these deep dopants would remove the aforementioned limitations in applications of this system. Because the origin of the zero-*B* dip is a coherent superposition of the spin states (further details can be found in Supplementary Note 12) [28, 29] S/Zn-codoped TFETs can be considered quantum functional devices.

## Methods

The device manufacturing process involved the formation of N+ and P+ electrodes, followed by the introduction of S and/or Zn dopants over the entire surface of the wafer, including the part that would later become the TFET channel. The gate stack was formed after this. Electron beam lithography was used to pattern the source, drain, and gate electrodes (Supplementary Note 3 presents further details). Four different (co-) doping conditions for deep dopants were adopted (only S, only Zn, S and Zn, and no deep dopant). The first two conditions were used for single-charge tunneling through a single species of deep dopant, the third condition was used for PSB for temperatures up to room temperature, and the fourth condition was used to confirm that the phenomenon described above does not occur in the absence of the deep dopants. The TFET channel must be sufficiently short to observe tunneling through deep dopant levels[44]. To realize a short channel within the scope of our process technology, we adopted a structure wherein part of the source/drain electrode overlapped the gate electrode, as shown in Fig. 1a, defining an effective channel length, $L_{eff}$. For the 13 devices introduced in this study, $L_{eff}$ was designed to range from 14 to 24 nm (Supplementary Table 1).

The transport characteristics of the device were evaluated following the standard three-terminal transistor evaluation procedure. A source–drain voltage $V_{SD}$ was applied to the P+ electrode of the device, a gate voltage $V_G$ was applied to the gate electrode, and the source–drain current $I_{SD}$ at the N+ electrode was measured. For magnetic field *B*, a superconducting magnet was used for measurements at temperatures below 250 K, and a solenoid or split electromagnet was used at room temperature (300 K). Furthermore, *B* was applied along the in-plane source–drain direction (crystal orientation <110>). For the alternating magnetic field at low-*B* resonance, an alternating current was applied to a coil (five turns of diameter 1 mm) installed 1 mm above the device substrate. To realize a high-*B*, microwaves were used to apply an alternating current to the backplate of the device substrate[20].

## Acknowledgments

We thank E. Abe, M. Kawamura, J. Yoneda, and T. Tanamoto for their discussions. This work was partly supported by JST CREST (Grant No. JPMJCR1871), MEXT Quantum Leap Flagship Program (Q-LEAP) (Grant No. JPMXS0118069228), and by JSPS KAKENHI (Grant No. 15H05867).

## Author contributions

Conceptualization: KO; Methodology: YB, KK, SI, HK, SMu, TM; Investigation: YB, SMo, KI, KO; Visualization: YB, KO; Funding acquisition: KI, TM, KO; Project administration: KO; Supervision: TM, KO; Writing – original draft: KO.

## Competing interests

The authors declare no competing interests.

## Data availability

The data supporting the findings of this study are available from the corresponding authors upon reasonable request.

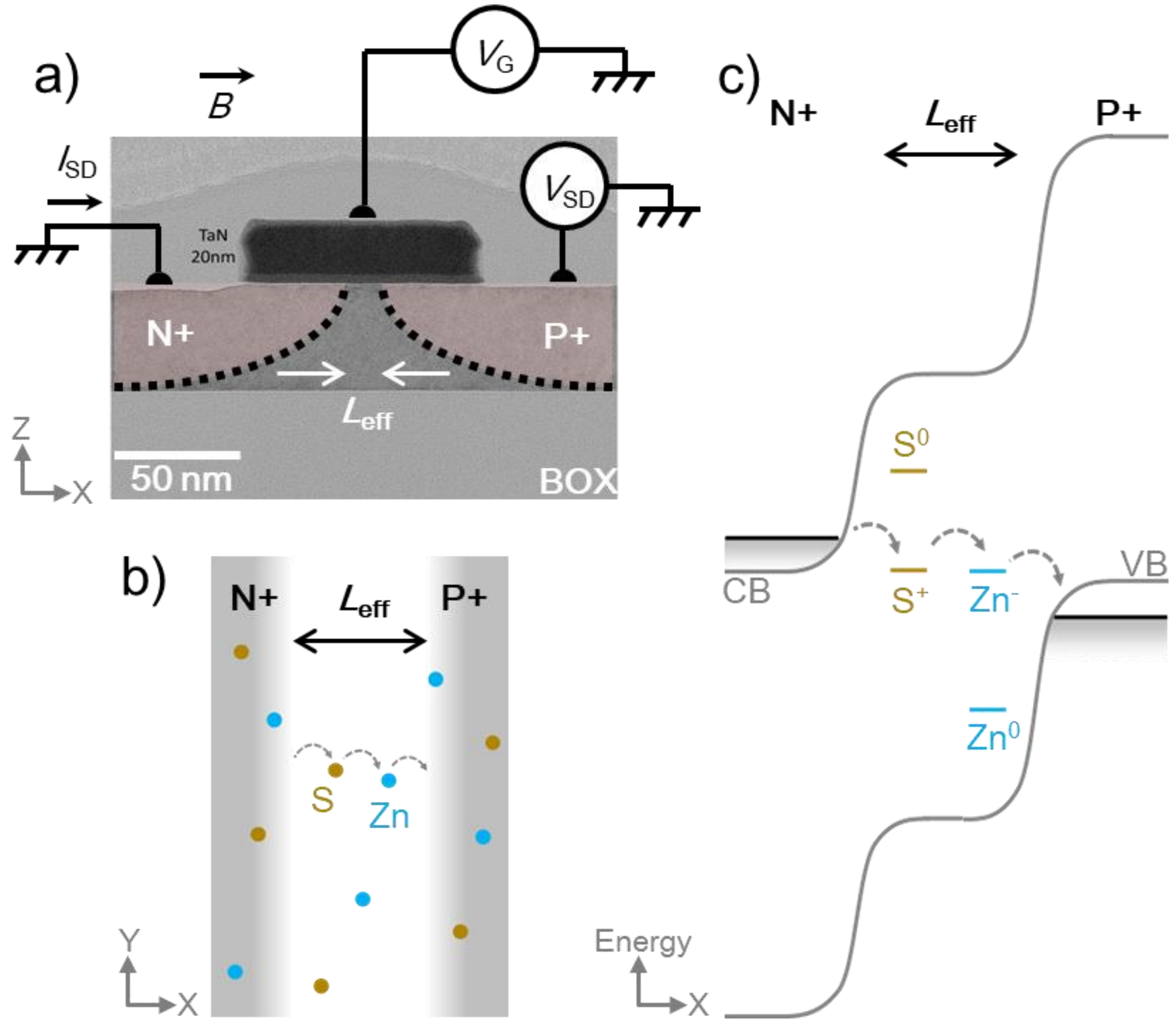

**Fig. 1. Tunnel field effect transistor (TFET) and deep dopants. a** Schematic of the device based on a cross-sectional transmission electron microscopy image. The gate overlaps the N+/P+ electrodes and defines the effective channel length $L_{eff}$. BOX: buried oxide. $V_{SD}$, $V_G$, and $I_{SD}$ contacts for transport characterization and the direction of $B$ are drawn. **b** S and Zn dopants in the channel and three-step tunneling of electrons from the N+ to the P+ electrode through S and Zn dopants in series. **c** The same three-step tunneling as that in the p-type intrinsic n-type (PIN) energy band diagram. It shows an $S^0$ level at 0.25 eV below the bottom of the conduction band (CB), an $S^+$ level at 0.5 eV below CB, and Zn dopants levels ($Zn^0$ level of 0.27 eV above the top of the valence band (VB) and $Zn^-$ level of 0.62 eV above the VB). These values are obtained using deep-level transient spectroscopy measurements of the S and Zn dopants introduced under the same conditions as that of the device fabrication[40]. In single charge transport, these are the levels with electron (hole) numbers of 1 and 2, respectively. Tunneling from the $S^+$ level to the $Zn^-$ level (or donor–acceptor recombination) is affected by the Pauli spin blockade (PSB). $V_{SD}$ source–drain voltage, $V_G$ gate voltage, $I_{SD}$ source–drain current, $B$ magnetic field.

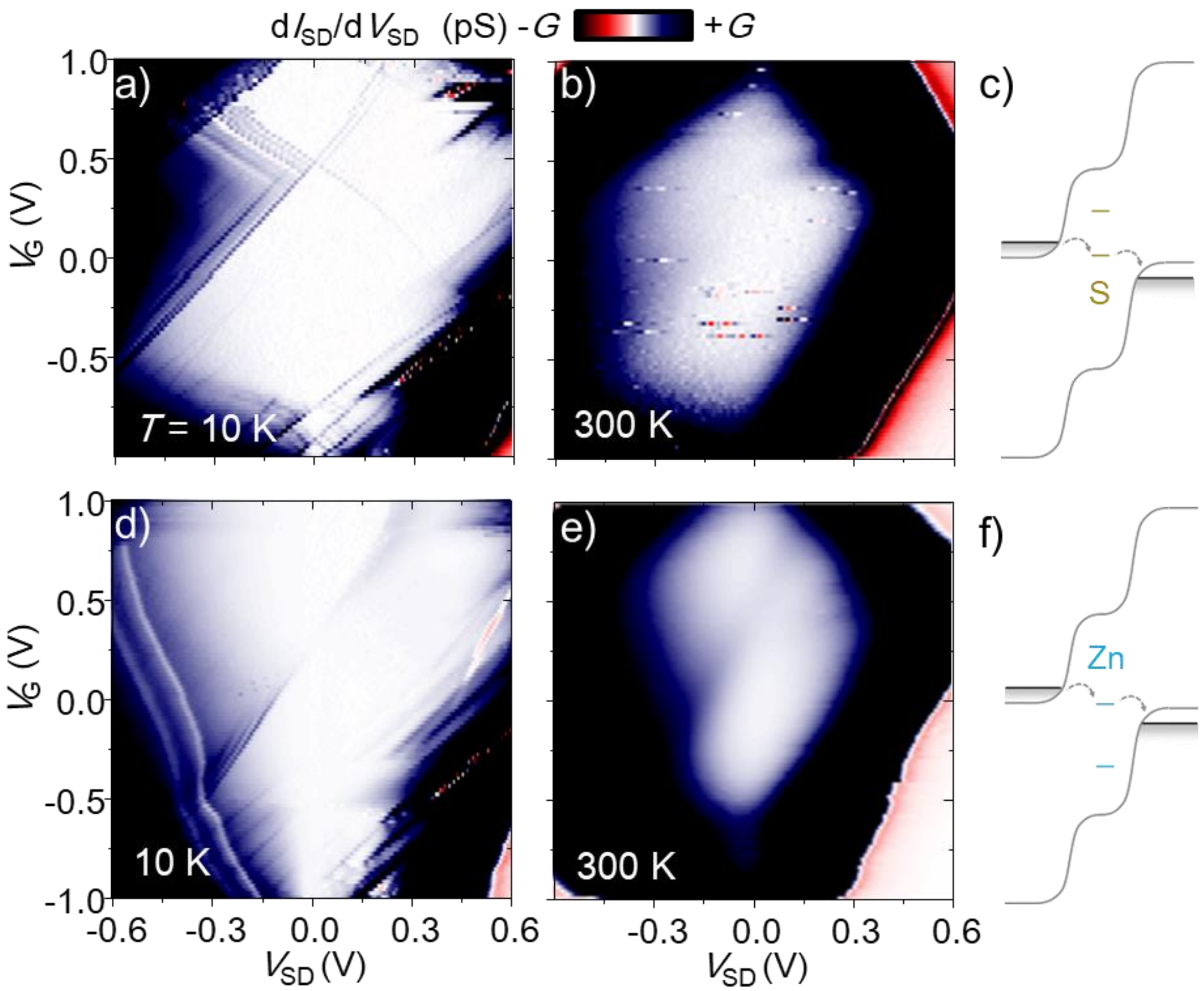

**Fig. 2. Single-charge tunneling in S (only)-doped, and Zn (only)-doped tunnel field effect transistor (TFET). a** Intensity plot of d$I_{SD}$/d$V_{SD}$ versus ($V_{SD}$, $V_G$) for a TFET with only S doped at $T$ = 10 K. **b** Similar plot at 300 K. **c** Energy band diagram showing the two-step tunneling transport through a single S dopant (two levels) existing in the I part (channel) of the p-type intrinsic n-type PIN structure. **d**, **e** Similar plots for a TFET with only Zn doped at temperatures of 10 and 300 K. **f** Schematic of single-charge transport through a single Zn dopant (two levels). The ranges of color intensity scale $G$ (pS) for plots a, b, d, and e, are 50, 50, 500, and 500, respectively. $V_{SD}$ source–drain voltage, $V_G$ gate voltage, $I_{SD}$ source–drain current, $T$ temperature.

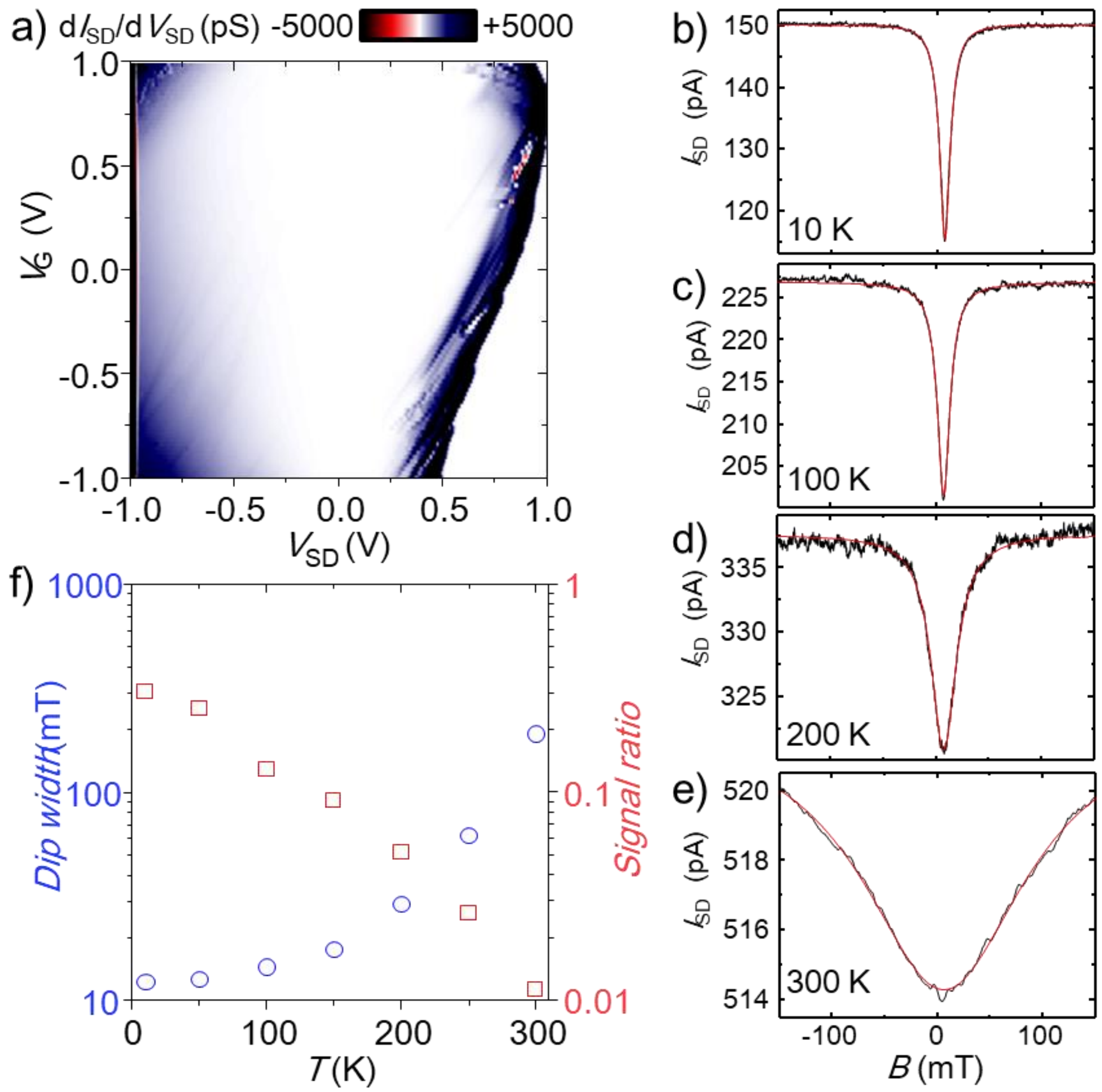

**Fig. 3. Zero-*B* structures up to room temperature in S/Zn-codoped tunnel field effect transistor (TFET). a** Intensity plot of $dI_{SD}/dV_{SD}$ versus ($V_{SD}$, $V_G$) at $T = 10$ K. **b-e** $B$ dependence of $I_{SD}$ observed for a specific ($V_{SD}$, $V_G$) (see Supplementary Note 5). $T = 10$, 100, 200, and 300 K for b-e, respectively. The solid lines represent Lorentzian fits. **f** Temperature dependence of the dip width (circles, left axis) and signal ratio (normalized change in $I_{SD}$ within the applied range of $B$, square, and right axis). $V_{SD}$ source–drain voltage, $V_G$ gate voltage, $I_{SD}$ source–drain current, $T$ temperature, $B$ magnetic field.

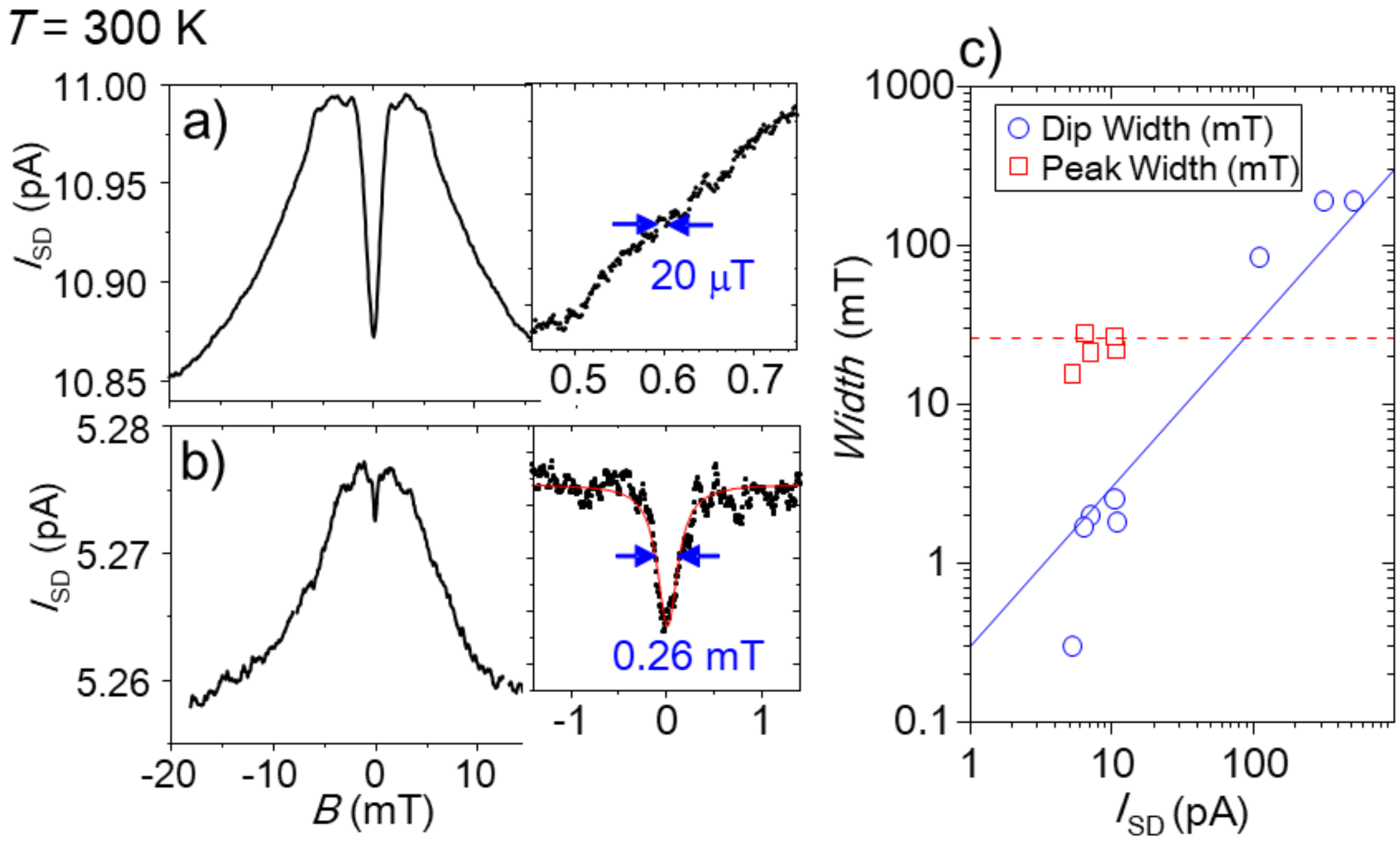

**Fig. 4**. **Room temperature zero-*B* structures in S/Zn-codoped tunnel field effect transistors (TFETs). a** *B* dependence of $I_{SD}$ for another S/Zn TFET. The inset presents the details at approximately 0.6 mT. **b** *B* dependence of $I_{SD}$ for another S/Zn TFET. The inset shows the details of a zero-*B* structure. The average results of 1000 repeated measurements are shown in a and b. **c** Summary of room-temperature zero-*B* structures. The dip widths obtained for the eight S/Zn-codoped TFETs (circles) and peak envelope widths obtained for the five S/Zn-codoped TFETs (squares) are plotted against their zero-*B* $I_{SD}$ values. The solid line represents the reference for linear correlation. The dashed line indicates the peak envelope width for the estimated nuclear fluctuation of 26 mT. $I_{SD}$ source–drain current, *T* temperature, *B* magnetic field.

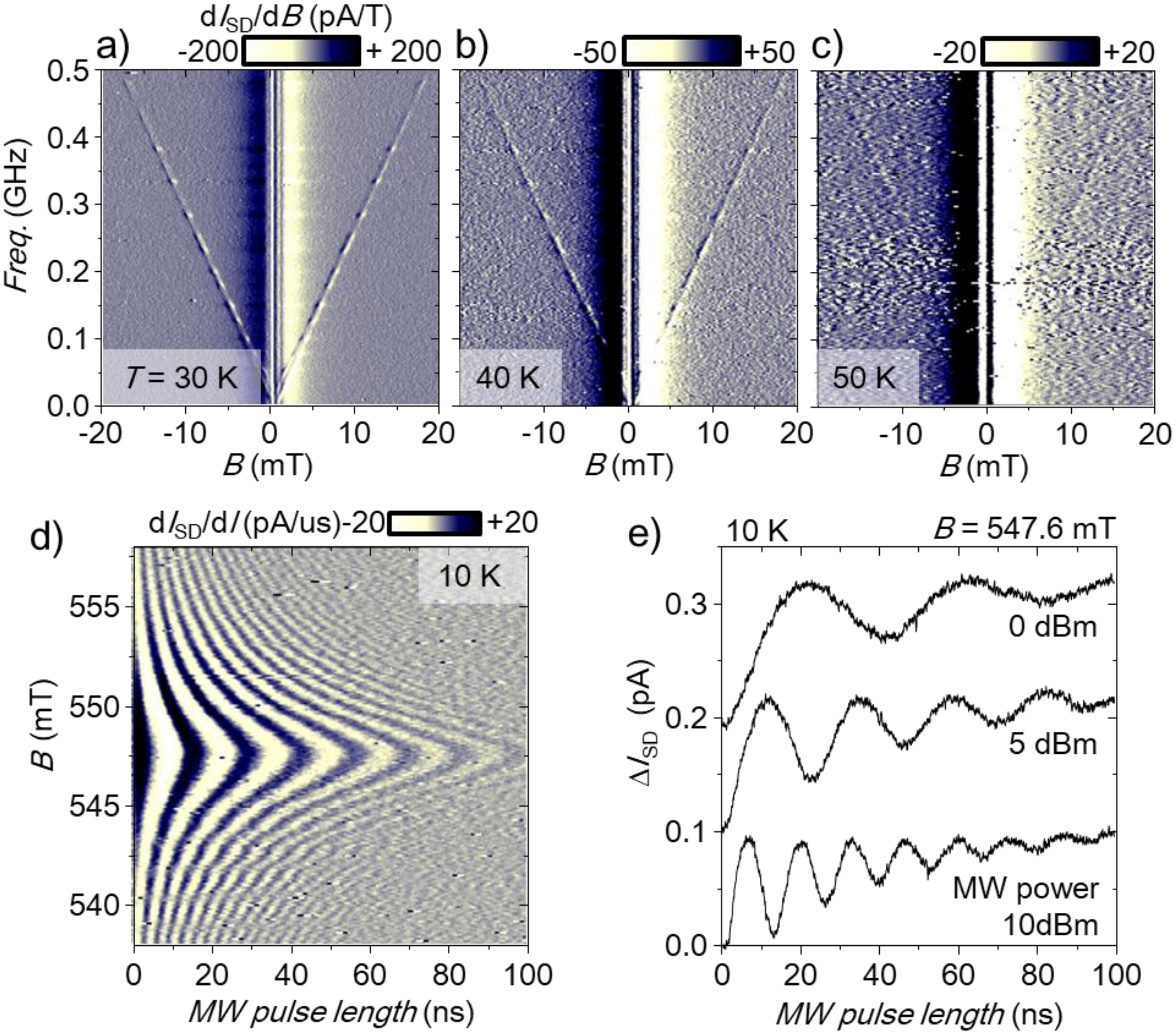

**Fig. 5. Magnetic resonance and Rabi oscillations in S/Zn-codoped tunnel field effect transistor (TFET). a-c** Subject to the settings of $V_{SD}$ and $V_G$ at which a zero-$B$ dip-peak structure is observed, $I_{SD}$ is measured against $B$ and the frequency (*Freq.*) of the alternating current magnetic field, and $dI_{SD}/dB$ is the plotted intensity. $T$ values for a-c are 30, 40, and 50 K, respectively. **d** Intensity plot of $dI_{SD}/dl$ when applying pulse-modulated microwaves (power 10 dBm, pulse period 500 ns, pulse length $l$ = 0–100 ns) at a frequency of 18.65 GHz, $T$ = 10 K. **e** Pulse-length dependence of the change of $I_{SD}$ ($\Delta I_{SD}$) at a resonant magnetic field 547.6 mT, measured at three microwave powers (10, 5, 0 dBm, respectively, offset by 0.1 pA), $T$ = 10 K. $V_{SD}$ source–drain voltage, $V_G$ gate voltage, $I_{SD}$ source–drain current, $T$ temperature, $B$ magnetic field, MW microwave.

# Supplementary information for

## Pauli spin blockade at room temperature in double-quantum-dot tunneling through individual deep dopants in silicon

Yoshisuke Ban, Kimihiko Kato, Shota Iizuka, Hiroshi Oka, Shigenori Murakami, Koji Ishibashi, Satoshi Moriyama, Takahiro Mori, and Keiji Ono

Corresponding author: k-ono@riken.jp

**This file includes:**

Supplementary Notes 1–12
Supplementary Figures 1–8
Supplementary Table 1

Supplementary Note 1. Infrared absorption and magnetic resonance of S and Zn dopants in Si

Although deep-level transient spectroscopy (DLTS) provides information on the dopant level for different charge states (charging energy), information on the excited states within the same charge state can be obtained via infrared absorption. According to[35], the gaps between the ground and lowest excited states for $S^+$ and $S^0$ are 0.43 and 0.28 eV, respectively. Both are approximately an order of magnitude larger than the thermal energy at room temperature (26 meV). The ground state and low-lying excited state are in the same *s* orbital, reflecting six valley degrees of freedom. The six symmetrical superposition states are the ground states. References[36, 37] indicate that the gap between the ground and excited states of $Zn^0$ is 0.30 eV. Although infrared absorption for $Zn^-$ has not been reported, the gap for $Zn^-$ is considered to be sufficiently larger than the thermal energy at room temperature. The infrared absorption experiments that yielded the aforementioned gap values were all conducted at low temperatures (1.5–10 K). Although gap values have not been measured at room temperature, it is reasonable to assume that there is no significant change in this value. Note that S and Zn in Si do not exhibit photoluminescence, except for complex dopants or dopants coupled with some type of defect center.

Magnetic resonance experiments provided information on the ground state manifolds of $S^+$ and $Zn^-$, particularly their spins. The $S^+$ ground state had a spin of 1/2 ($g$ factor = 2.0054)[38]. The ground-state manifold of $Zn^-$ comprised a 3/2 spin-orbit state ($\Gamma_8$) and a 1/2 spin state ($\Gamma_7$), which reflected the spin–orbit character of the valence band. The energy gap between $\Gamma_8$ and $\Gamma_7$ was 0.31 meV, which is considerably smaller than that for the shallow acceptor (22 meV) and that for the valence band (43 meV). The magnetic resonance spectrum was strongly nonlinear with respect to the magnetic field strength and depended on the direction of the magnetic field. For example, the g factors in a weak magnetic field along the applied direction [001] were 0.71, 2.2, and 0.73 for the $\Gamma_7$ ±1/2, $\Gamma_8$ ±3/2, and $\Gamma_8$ ±1/2 states, respectively[39].

Supplementary Note 2. S/Zn doping

Preliminary studies were conducted to dope S and Zn into Si[40]. Intended for doping into tunnel field effect transistors (TFETs), these dopants were introduced close to the surface (<10 nm) of the substrate via ion implantation. A millisecond post-annealing condition was adopted as the heat treatment to recover the implantation damage and achieve a dopant density profile similar to that in a previous study[20]. The dopant density profile was confirmed by secondary ion-mass spectroscopy[40]. Notably, owing to thermal treatment under a non-equilibrium state, this density exceeded the solid solubility limit of S and Zn dopants by more than two orders of magnitude[45]. The dopant levels of S and Zn were evaluated by DLTS and were nearly consistent with those of previous studies[31–34, 40]. Simultaneously, transmission electron microscopy images and photoluminescence analysis confirmed the absence of defects in Si crystal caused by ion implantation[40]. Note that the defects here are not the same as the deep dopants, and the two are different.

Supplementary Note 3. Details of the device manufacturing process

The TFETs, gated p-type intrinsic n-type (PIN) diodes, were fabricated using a process similar to that used for metal-oxide-semiconductor field-effect transistors. The process comprises device isolation, source and drain formation by ion implantation, their activation by rapid thermal annealing, gate stack and etching taking advantage of high-k/metal gate, electrode formation, and hydrogen sintering. The TFETs were fabricated on silicon-on-insulator wafers. The silicon-on-insulator thickness was approximately 50 nm, and the buried oxide thickness was 145 nm. The top wafer surface was (100) and the current flow direction was <110>. The P+ source electrode was formed using $BF_2$ ion implantation with an energy of 2.5 keV and a dose of $2 \times 10^{15}$ $cm^{-2}$. The N+ drain electrode was formed using As ion implantation with an energy of 5 keV and a dose of $2 \times 10^{15}$ $cm^{-2}$. Rapid thermal annealing was performed at 950 °C for 2 s, which intends to re-crystallize the source and drain region (amorphized by the ion implantation process) and to activate donor and acceptor dopants implanted in the source and drain. Subsequently, we introduced S and/or Zn deep dopants into the active region using ion implantation with an energy of 15 keV and a dose of $1 \times 10^{13}$ $cm^{-2}$. For efficient implantation, the mass selection slit of the ion implanter was set around the isotopes with the largest natural abundance ratio, namely $^{32}$S and $^{64}$Zn. Flash lamp (millisecond) annealing was conducted at 1200 °C for 1.5 ms (after a stand-by temperature of 800 °C for 3 min) to activate deep dopants. Finally, the gate was produced using high-k/metal gate technology. Interfacial $SiO_2$ with a thickness of approximately 1 nm was chemically formed, and $HfO_2$ (2.4 nm) was deposited onto it using atomic layer deposition. A TaN gate with a thickness of 20 nm was formed by sputtering. The equivalent oxide thickness of the gate insulator was estimated to be approximately 1.5 nm based on capacitance–voltage measurements.

Although $L_{\mathrm{eff}}$ varies owing to factors such as the mask alignment accuracy in the lithography process, we believe that the relative relationship between the $L_{\mathrm{eff}}$ values for different TFETs is still a reliable indicator for the effective channel length.

## Supplementary Note 4. Supplemental information for Fig. 2

As shown in Fig. 2a (and to a certain extent in Fig. 2d), several Coulomb diamonds were observed at $V_G$ values close to -1.0 V, exhibiting much smaller charging energies compared with the diamonds at more positive $V_G$ values. The presence of these small diamonds could be attributed to the shallow dopants in the channel near the N+ electrode, because the channel becomes almost p-type and the transport of the TFET is dominated by PIN tunneling through these shallow dopants at $V_G$ values close to -1V. Supplementary Figure 1 shows a band diagram for $V_G \sim$ -1V. By further increasing negative $V_G$ values, the channel became entirely p-type, and a band-to-band tunneling (Zener tunneling) between the channel and the N+ electrodes is responsible for the on-current of the TFET. Similarly, for large positive $V_G$ values, a band-to-band tunneling between the P+ electrodes and the channel (entirely n-type) is responsible for the on-current of the TFET.

## Supplementary Note 5. Supplemental information for Fig. 3

Supplementary Figure 2a shows the detailed temperature dependence of the zero-$B$ dip structure. The background $I_{SD}$ increased with increasing temperatures, as shown in Supplementary Figure 2b. Supplementary Figure 2c shows the temperature dependence of optimal $V_G$. The optimum $V_G$, at which these dip structures could be observed most clearly, changed with temperature. At temperatures above 150 K, the optimal $V_G$ gradually shifted in the positive direction. This change can be attributed to the temperature dependence of the gate metal-oxide semiconductor capacitance. The same transport conditions were maintained by shifting $V_G$, thus compensating for the change in capacitance. A similar $V_G$ shift was also observed in the temperature dependence of the Coulomb oscillation in Fig. 2(c) of [20]. The optimal $V_{SD}$ was maintained at 0.5 V. Supplementary Figure 2d shows the $V_G$ dependence of the zero-$B$ dips measured at 150 K, showing the optimum $V_G$ for this temperature. There is a single optimum $V_G$ value and the contrast of the dip decays monotonically in either $V_G$ direction. At this optimal $V_G$, the deep dopant levels that contribute to PSB are thought to be the most thermally stable (as also shown in Fig. 1c).

Supplementary Note 6. Two-step single charge tunneling in S/Zn codoped device

Supplementary Figures 3a and 3b show the results of Coulomb diamond measurement similar to those in Figs. 3a, 3b, 3d, and 3e for a device with S/Zn-codoped TFET. At approximately $V_G$ = 0.25 V, where zero-bias conductance was restored, charge transport with two-step tunneling was dominant. These results indicate that tunneling transport occurred through a single deep dopant level in this device. Transport models are schematically depicted in Supplementary Figure 3c.

Supplementary Note 7. Supplemental information for Fig. 4

Supplementary Figure 4 shows the room-temperature zero-*B* structures of five other S/Zn-codoped TFETs. The plot in Fig. 4c was constructed by combining the results of the five S/Zn-codoped TFETs and three S/Zn-codoped TFETs discussed in the main text. The average results of 1000 repeated measurements are shown.

Supplementary Note 8. Random fluctuation of nuclear spins near deep dopants

According to [27], the full-width-at-half-maximum of the zero-*B* peak envelope depends on the magnitude of the nuclear fluctuations. In our case, the peak width was determined from the natural abundance ratio of approximately 5 % $^{29}$Si, hyperfine coupling constant, and the number of $^{29}$Si atoms within the dot.

In [43], a quantum dot made of natural Si was estimated to contain $10^5$ Si atoms (5000 $^{29}$Si) and exhibit a hyperfine coupling constant of 1.85 mT. The typical confinement of the quantum dot was parabolic (5 meV) for in-plane directions and rectangular (100 meV) for the perpendicular to plane direction, and that of an S dopant was Coulombic (300 meV) for all directions. Because the approximate size of the dot's or dopant's wave function is inversely proportional to the square root of the confinement energy, the ratio in their volume was ~ $(300/5)^{1/2} \times (300/5)^{1/2} \times (300/100)^{1/2} \sim 100$. From this volume ratio, for S dopants, the total number of Si atoms was $10^5/100 = 1000$, number of $^{29}$Si atoms was $1000 \times 0.05 = 50$, hyperfine coupling constant was $1.85 \times 100 = 185$ mT, and the nuclear fluctuation was $185/(50)^{1/2} \sim 26$ mT.

Another approach for estimating the nuclear fluctuations for S in Si involved the conduction of infrared absorption experiments in S-doped Si. In [35], the infrared absorption spectrum of $^{32}$S in natural Si ($^{Nat}$Si:$^{32}$S) was compared with that of $^{28}$Si-enriched Si ($^{28}$Si:$^{32}$S). Almost all spectral peaks had widths of 20 μeV for $^{Nat}$Si, whereas for $^{28}$Si, they were an order of magnitude smaller. This decrease in linewidth was attributed to $^{28}$Si suppressing the nuclear mass variation in $^{Nat}$Si. However, this decrease should also include the removal of nuclear fluctuations owing to the removal of $^{29}$Si. Therefore, the *B* value of 170 mT, converted from a linewidth of 20 μeV with a *g*-factor of 2.0, provided the upper bounds of the nuclear fluctuation at the S site.

Supplementary Note 9. Supplemental information for Fig. 5

As supplementary data for Fig. 5a-c, Supplementary Figure 5a shows the dependence of $I_{SD}$ on *B* at an alternating current magnetic field frequency of 382.5 MHz at *T* = 30, 40, and 50 K. The zero-*B* dip, peak envelope, and magnetic resonance peaks (*B* = ±14 mT) were observed. Supplementary Figure 5b represents supplementary data for Fig. 5d. Herein, $dI_{SD}/dB$ is the intensity plotted with respect to the direct current magnetic field *B* and microwave frequency (*Freq.*). A straight diagonal line owing to magnetic resonance is observed in certain parts.

Supplementary Note 10. Disappearance of magnetic resonance response

The disappearance of the magnetic resonance can be explained by assuming that the three coherent spin-flip tunneling rates ($11T_+$ to $02S$), ($11T_0$ to $02S$), and ($11T_-$ to $02S$) become equal. In this case, the magnetic resonance transition between the triplet states did not cause a change in $I_{SD}$. Moriyama et al.[46] evaluated the coherent spin-flip tunnel rates for a *02-11* charge transition (reversal of charge transition in PSB), and the coherent spin-flip tunnel rate was found to be comparable to the spin conservation tunnel rate. The application of this method to an S/Zn-codoped TFET can verify the above hypothesis.

If the spin 1/2 states ($\Gamma_7$) are the only states that $Zn^-$ can accept, the electron–hole recombination with the $S^+$ (spin 1/2) state should have spin 0 following recombination, which is typical of PSB. However, as the $Zn^-$ ground state manifold also possesses spin-orbit 3/2 states ($\Gamma_8$), must be considered. If the $Zn^-$ ground state manifold before recombination is $\Gamma_8$, the total angular momentum, including $S^+$, is a composite of 1/2 and 3/2; this composite state cannot have a total angular momentum of 0 after recombination. This is a type of PSB. Note that this PSB cannot be lifted even if the 1/2 electron spin of $S^+$ is rotated by magnetic resonance. Similarly, the PSB cannot be lifted through a magnetic resonance transition between two $\Gamma_7$ states or four $\Gamma_8$ states. This can be related to the disappearance of magnetic resonance. It is noted that a transition between $\Gamma_7$ and $\Gamma_8$ states was observed owing to electric dipole resonance[39]. This suggests that our PSB may be lifted using this $\Gamma_7$–$\Gamma_8$ transition together with the standard rotation of electron spin 1/2 of $S^+$.

Supplementary Note 11. Estimation of noise spectral density

Time-series data are needed to discuss the noise spectral density, which constitutes a future topic of research work. Herein, we evaluate the noise spectral density using pseudo time-series data corresponding to the measurements in the inset of Fig. 4a. One trace of the raw data in Fig. 4a (i.e., before the 1000th averaging cycle) is shown in Supplementary Figure 6a. The magnetic field was swept from -0.3 mT to1.4 mT over approximately 75 s, and 1000 current data points were acquired. The magnetic field dependence of this current was fitted with a Lorentzian. We used the standard residuals of the Lorentzian fit as a function of the measurement time as the pseudo time-series data. The noise spectral density obtained with the pseudo time-series data is shown in Supplementary Figure 6b. The current noise was converted to magnetic field noise using the slope of the plot in the inset of Fig. 4a.

Supplementary Note 12. Coherent population trapping and PSB

The zero-*B* dip structure is explained by PSB theory under strong SOI conditions[28, 29]. In this section, we provide a brief explanation of coherent population trapping (CPT) in a lambda-type quantum three-level system and compare it with that of PSB[47]. The goal is to facilitate non-experts in this field in obtaining a perspective of the physics governing the production of zero-*B* dip structures. We also discuss the possibility that the physics involved in PSB may be enriched by importing related physics concepts from the CPT research field.

The lambda system considered here was the minimum simple system necessary to understand PSB under SOI. As shown in Supplementary Figure 7a, it comprised two degenerate ground states “1” and “2” and one excited state “3”; the ground and excited states are separated by $\Delta$. Incoherent relaxation occurs from the excited state to the ground state at a rate $\Gamma$. There exists a coherent transition $\Omega_1$ between levels 1 and 3 and a coherent transition $\Omega_2$ between levels 2 and 3.

The Hamiltonian is expressed as

$$H = \begin{bmatrix} 0 & 0 & \Omega_1 \\ 0 & 0 & \Omega_2 \\ \Omega_1^* & \Omega_2^* & \Delta \end{bmatrix},$$

and its time evolution is expressed as the density matrix rate equation

$$\dot{\rho} = -i\left[\frac{H}{\hbar}, \rho\right] + \mathrm{L}[\rho].$$

The second term is the rate equation, which is expressed as

$$\mathrm{L}[\rho] = \begin{bmatrix} \Gamma\rho_{11} & 0 & 0 \\ 0 & \Gamma\rho_{22} & 0 \\ 0 & 0 & -2\Gamma\rho_{33} \end{bmatrix},$$

and describes the incoherent time evolution. Herein, we converted bases 1 and 2 into another set of bases. The bases in the subspace spanned by state vectors 1 and 2 were rotated, and new bases, 1' and 2', were set. At appropriate rotation angles, the new bases 1', 2', and 3 had zero coherent transitions between 2' and 3, as shown in Supplementary Figure 7b.

This is equivalent to using a rotation matrix

$$U = \begin{bmatrix} \cos\theta & -\sin\theta & 0 \\ \sin\theta & \cos\theta & 0 \\ 0 & 0 & 1 \end{bmatrix}.$$

Furthermore,

$$U^*HU = \begin{bmatrix} \cos\theta & -\sin\theta & 0 \\ \sin\theta & \cos\theta & 0 \\ 0 & 0 & 1 \end{bmatrix} \begin{bmatrix} 0 & 0 & \Omega_1 \\ 0 & 0 & \Omega_2 \\ \Omega_1^* & \Omega_2^* & \Delta \end{bmatrix} \begin{bmatrix} \cos\theta & \sin\theta & 0 \\ -\sin\theta & \cos\theta & 0 \\ 0 & 0 & 1 \end{bmatrix}$$

$$= \begin{bmatrix} 0 & 0 & \sqrt{\Omega_1^2 + \Omega_2^2} \\ 0 & 0 & 0 \\ \sqrt{\Omega_1^2 + \Omega_2^2} & 0 & \Delta \end{bmatrix} \text{for } \theta = \tan^{-1}\frac{\Omega_2}{\Omega_1}.$$

A steady-state solution is

$$\rho = \begin{bmatrix} 0 & 0 & 0 \\ 0 & 1 & 0 \\ 0 & 0 & 0 \end{bmatrix},$$

and the occupation rate of state 2' is 1. State 2' is a coherent superposition of states 1 and 2. This is called CPT because the state of the lambda system is trapped in a coherent superposition state.

One feature of CPT is its sensitivity to the degeneracy of states 1 and 2. In case of a small energy difference $\delta$ between states 1 and 2, as shown in Supplementary Figure 7c, the results of the same rotation as above are as follows:

$$U^*HU = \begin{bmatrix} \cos\theta & -\sin\theta & 0 \\ \sin\theta & \cos\theta & 0 \\ 0 & 0 & 1 \end{bmatrix} \begin{bmatrix} 0 & 0 & \Omega_1 \\ 0 & \delta & \Omega_2 \\ \Omega_1^* & \Omega_2^* & \Delta \end{bmatrix} \begin{bmatrix} \cos\theta & \sin\theta & 0 \\ -\sin\theta & \cos\theta & 0 \\ 0 & 0 & 1 \end{bmatrix}$$

$$= \begin{bmatrix} \delta\sin^2\theta & \delta\sin\theta\cos\theta & \sqrt{\Omega_1^2 + \Omega_2^2} \\ \delta\sin\theta\cos\theta & \delta\cos^2\theta & 0 \\ \sqrt{\Omega_1^2 + \Omega_2^2} & 0 & \Delta \end{bmatrix}.$$

Thus, a non-zero coherent transition remains between 1' and 2', as shown in Supplementary Figure 7d. Therefore, the CPT to 2' is lifted.

PSB can be understood as a lambda system. Supplementary Figure 8a shows a schematic of the three charge-spin states, *S02*, *S11*, and *T11*, in the PSB and the transitions between them.

Herein, following the standard PSB system for double dots, the numbers of electrons 1 and 2 were used for the carrier occupancy of the Zn site instead of the numbers of holes 0 and 1. The $\Gamma_8$ state (spin-orbit 3/2 state) of Zn was not considered. The solid arrow labeled with $t_S$ represents a coherent tunneling process between dopants without spin flip, and the solid arrow labeled with $t_T$ represents coherent spin-flip tunneling owing to SOI. In *S02*, the tunneling process from the Zn site to the drain and the subsequent tunneling from the source to the S site can be treated as an incoherent relaxation from *S02* to *S11* (or *T11*) via the *01* state, as indicated by the dashed arrows. Note that *S02* does not have a higher energy than *S11* or *T11*. *S02* is presented at the top of the panel for convenience in describing the incoherent relaxations. The threefold degeneracy of *T11* has not been considered.

As discussed above, the *S11* and *T11* states were rotated to create the new states *S'11* and *T'11*. For an appropriate rotation angle, as shown in Supplementary Figure 8b, the coherent transition between *T'11* and *S02* disappeared, and CPT occurred at *T'11*. *S'11* and *T'11* were the coherent superpositions of the spin singlet and triplet states, respectively. In contrast to the (anti) parallel spins for *T11* (*S11*) in Supplementary Figure 8a, the spin states for *T'11* and *S'11* are presented to be noncollinear, as shown in Supplementary Figure 8b.

In the magnetic field, both $T_-11$ and $T_+11$ states were Zeeman-shifted with respect to $T'_011$, and the three-fold degeneracy of *T'11* was lifted. The CPT for these two states was then lifted, leaving $T_0'11$ as the only CPT state, as shown in Supplementary Figure 8c.

In summary, there were three CPT states for zero *B* and one CPT state for nonzero *B*. If we consider a realistic process that provides a finite lifetime for the PSB, such as co-tunneling, the PSB is most effective at zero *B*. Thus, the PSB leakage current has a zero-*B* dip structure.

CPT was proposed in 1976[48] and has been studied since then in various fields, such as atomic gas, trapped ions, superconducting circuits, and diamond nitrogen-vacancy centers. The history of CPT is extensive, and the physical systems covered are broader than those of PSB. The possibility of importing various related physics from the CPT field into the PSB field is compelling. For example, when both transition rates $\Omega_1$ and $\Omega_2$ are simultaneously on-off modulated in a time scale of the coherent time, the occupancy of state 2' oscillates sensitively against $\delta$. This is known as Raman–Ramsay CPT[49]. Furthermore, in a lambda system coupled with nuclear spin degrees of freedom, the dip owing to CPT against $\delta$ is multiplied by the number of nuclear spin levels[50].

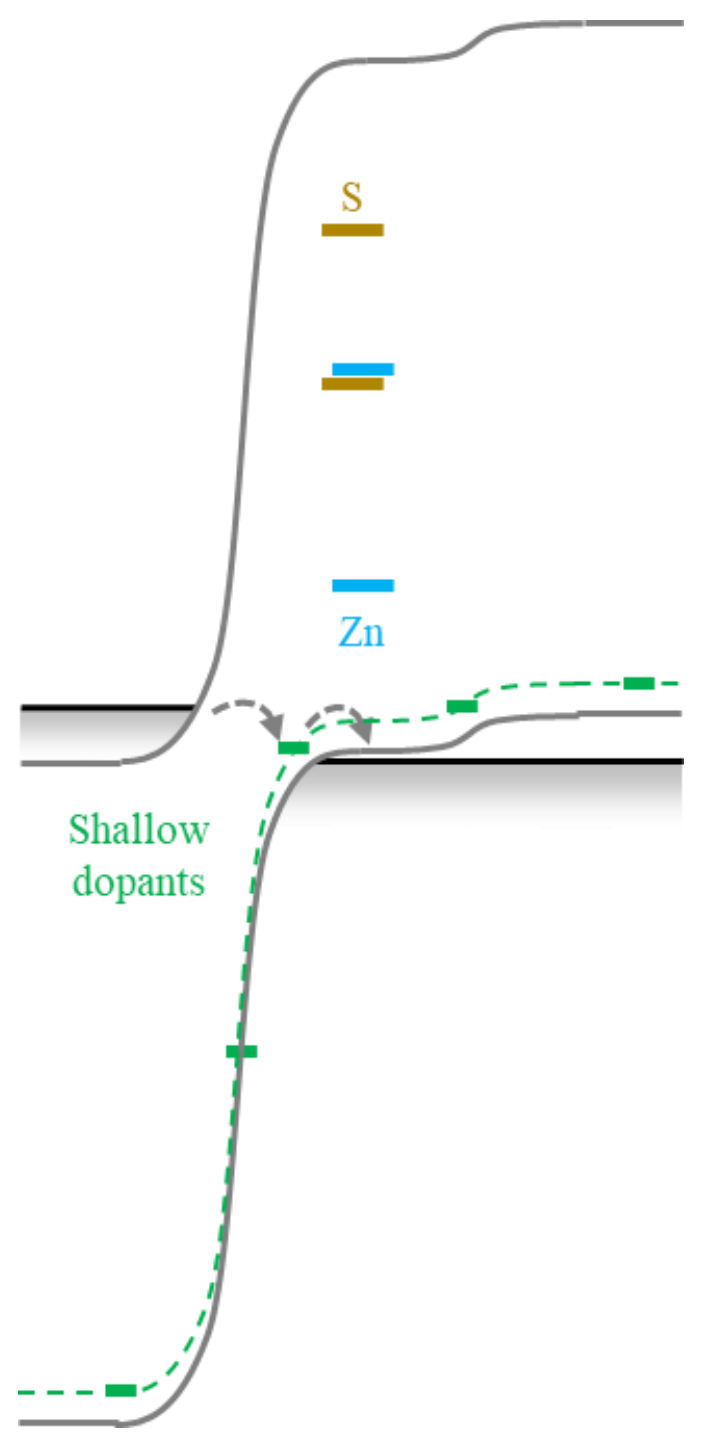

**Supplementary Figure 1.**

Energy band diagram for $V_G \sim -1$ V.

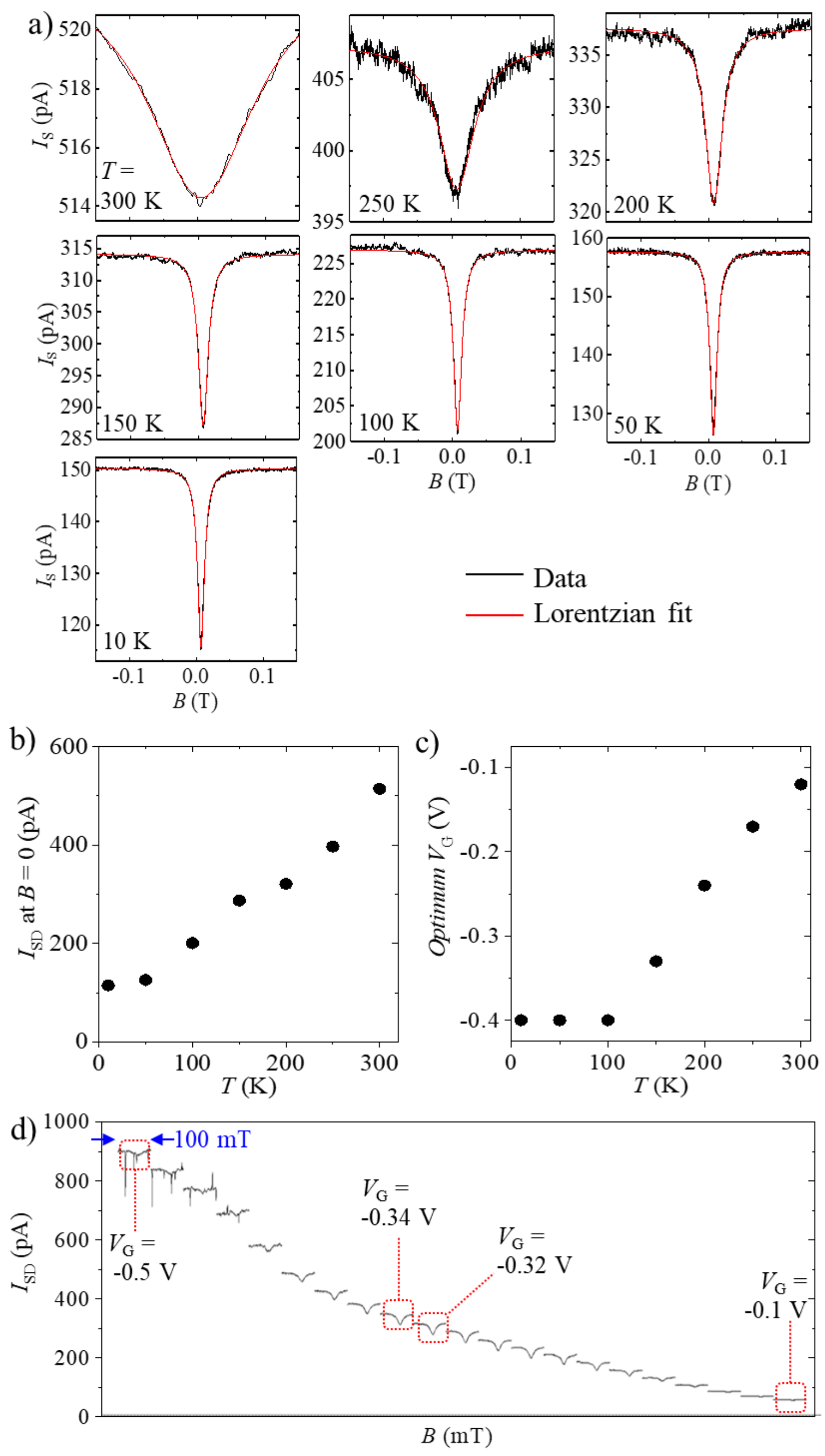

**Supplementary Figure 2.**

**a** Detailed temperature dependence of zero-$B$ dip structures, temperature dependence of zero-$B$ $I_{SD}$ (**b**), and optimum $V_G$ (**c**). **d** $V_G$ dependence of the zero-$B$ dips measured at 150 K. Each curve was laterally shifted by 100 mT ($B$ was swept from -50 to 50 mT), and measured at $V_G$ = -0.5 V (left end) to -0.1 V (right end) with a 0.02 V step, while keeping $V_{SD}$ = 0.5 V. The optimum $V_G$ value can be recognized at approximately -0.33 V.

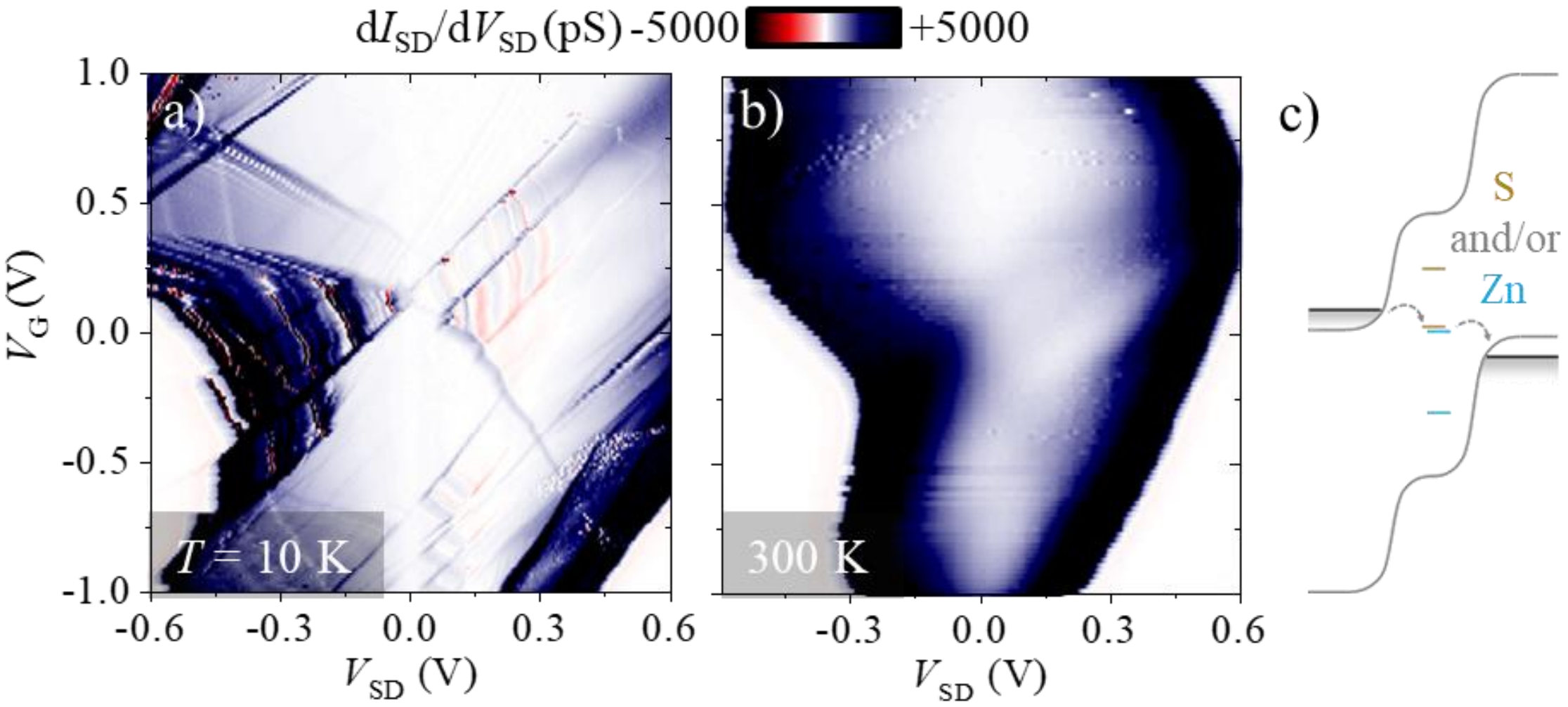

**Supplementary Figure 3.**

Two-step single-charge transport in S/Zn-codoped TFET. **a** Intensity plot of differential conductivity d$I_{SD}$/d$V_{SD}$ versus ($V_{SD}$, $V_G$) at $T$ = 10 K. **b** Same plot at a temperature of 300 K. **c** Schematic of single-electron transport via S and/or Zn dopant.

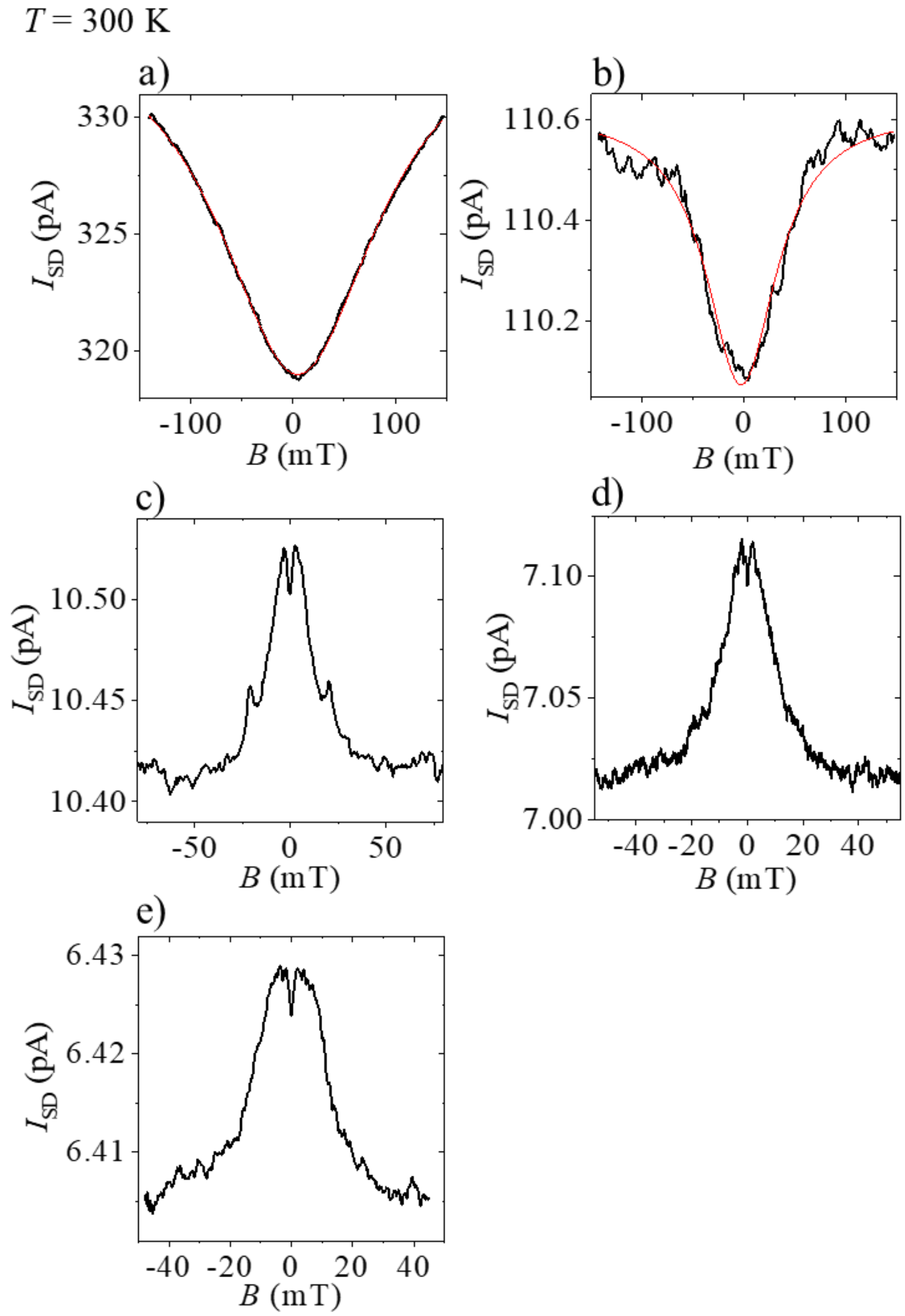

**Supplementary Figure 4.**

**a-e** Room temperature zero-*B* structures obtained for five other S/Zn-codoped TFETs.

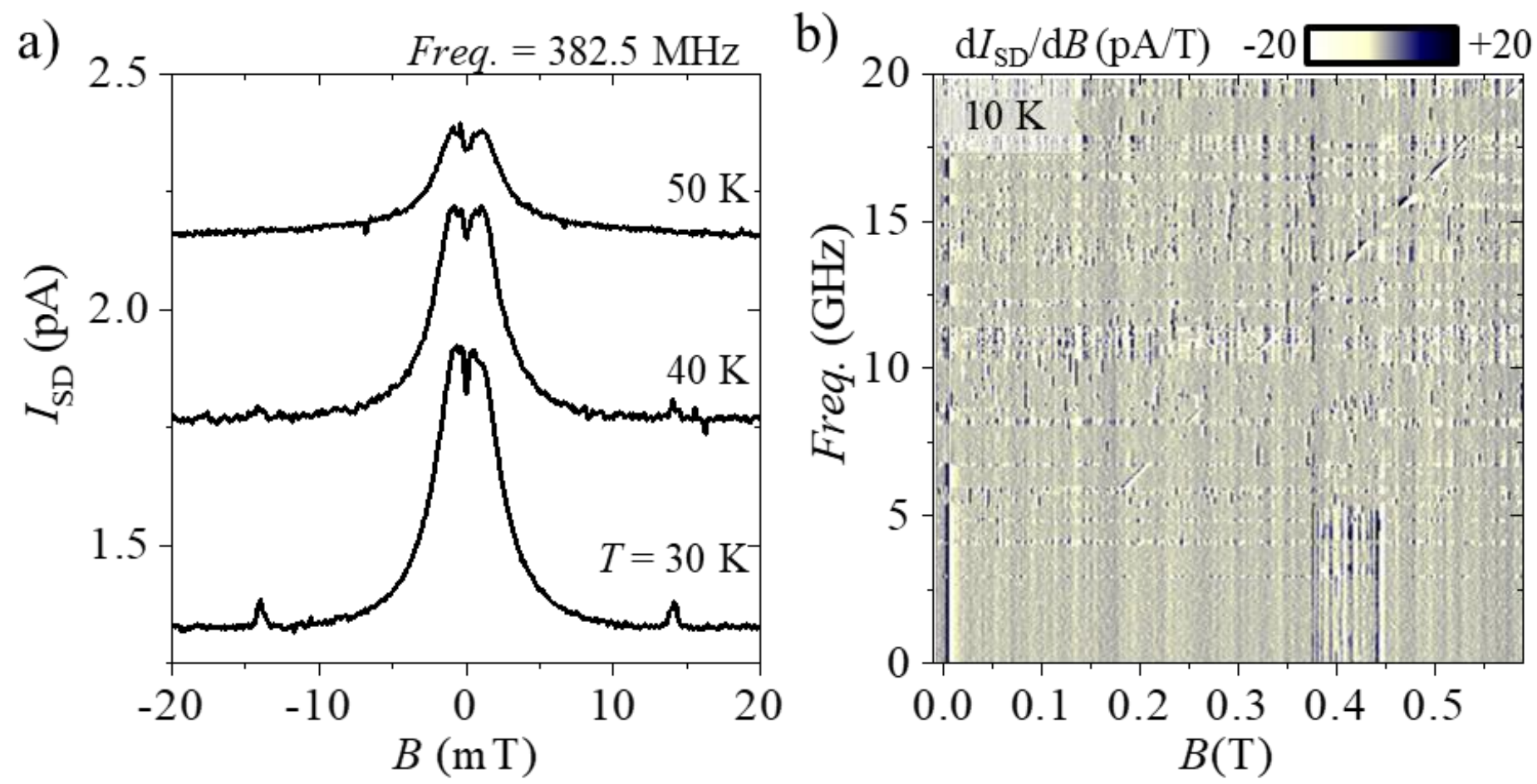

**Supplementary Figure 5.**

**a** Dependence of $I_{SD}$ on $B$ at temperatures of 30, 40, and 50 K with alternating current magnetic field frequency of 382.5 MHz for the device in Figs. 5a-c. **b** Color intensity plot of d$I_{SD}$/d$B$ versus $B$ and microwave frequency (*Freq.*) for the device in Figs. 5d and 5e.

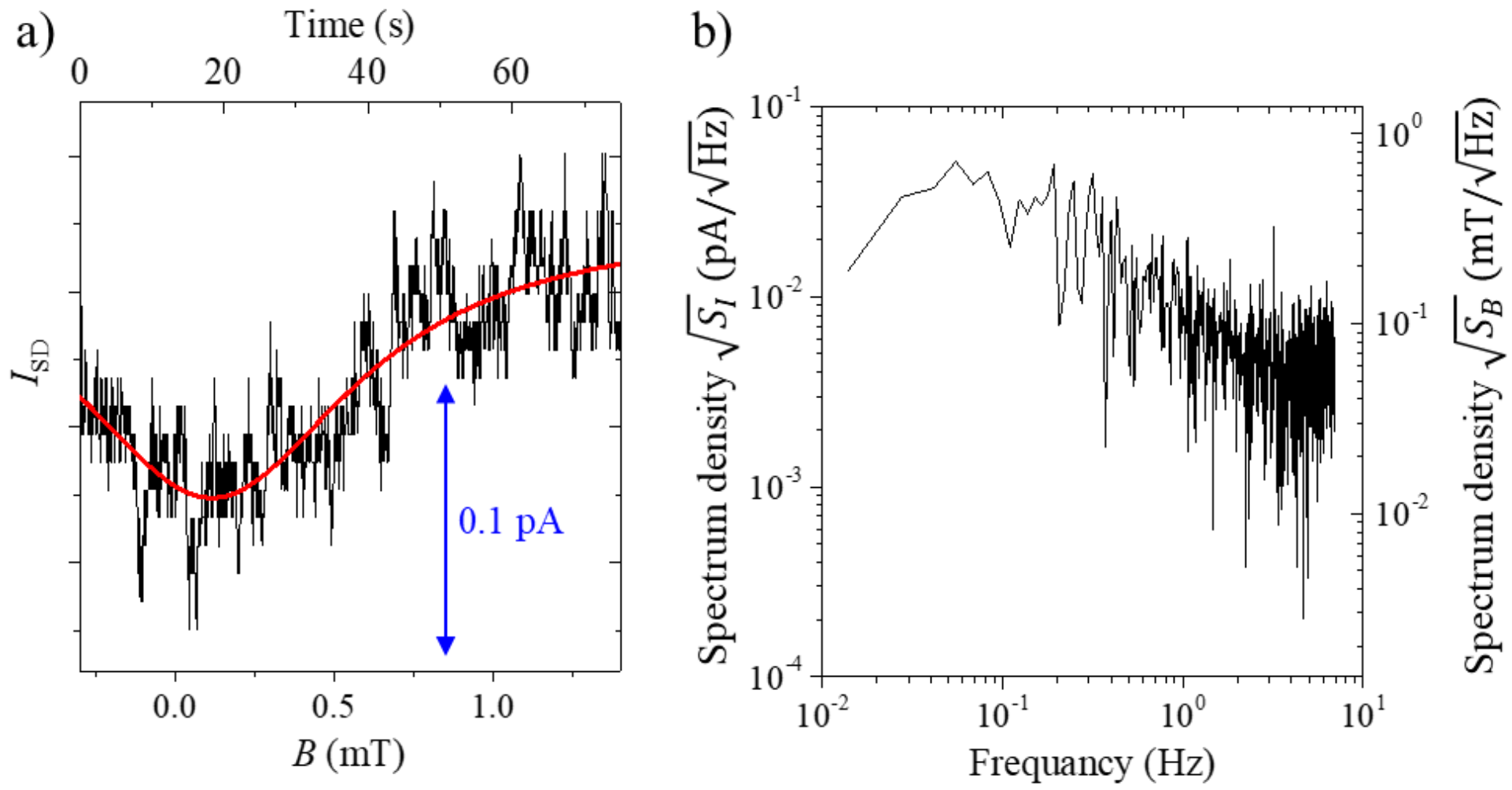

**Supplementary Figure 6.**

**a** Single trace of $I_{SD}$ vs. $B$ for the data of the inset in Fig. 4a. The upper axis shows the measurement times. The red line lines represent the Lorentzian fits. **b** Current noise spectral density obtained from the pseudo time-series data. The axis on the right shows the magnetic field noise spectral density that is converted from the current noise spectral density using the slope of the plot in the inset of Fig. 4a.

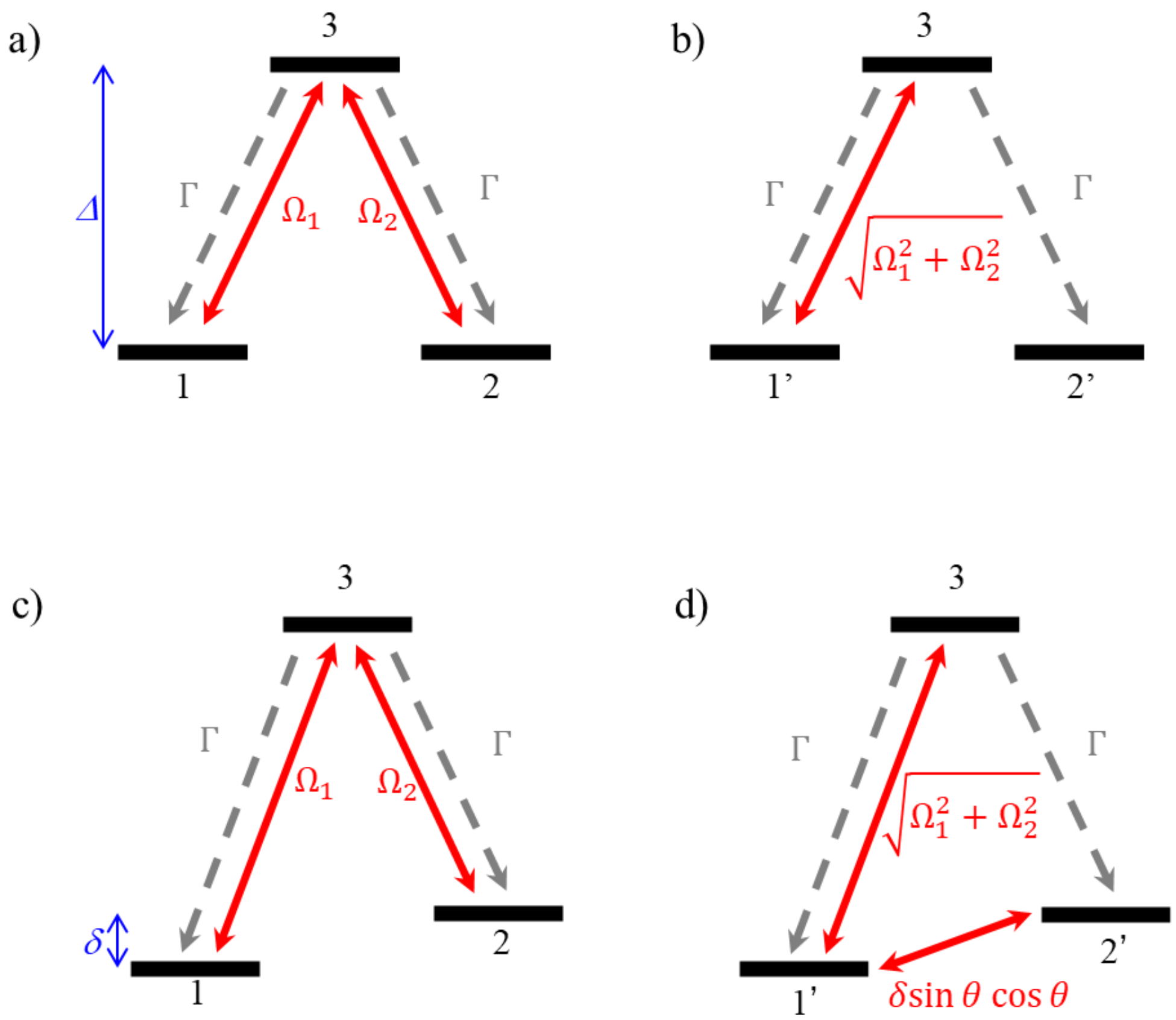

**Supplementary Figure 7.**

Lambda-type quantum three-level system. **a** Lambda system comprising two degenerate ground levels 1 and 2 and one excited level 3. **b** Bases 1 and 2 are rotated in the sub-Hilbert space and become 1' and 2'. The transition between 2' and 3 vanishes, resulting in a coherent population trapping (CPT) at 2' for suitable rotations. **c** Cases of an energy difference $\delta$ between levels 1 and 2. **d** Coherent transition remains between 1' and 2', and CPT does not occur.

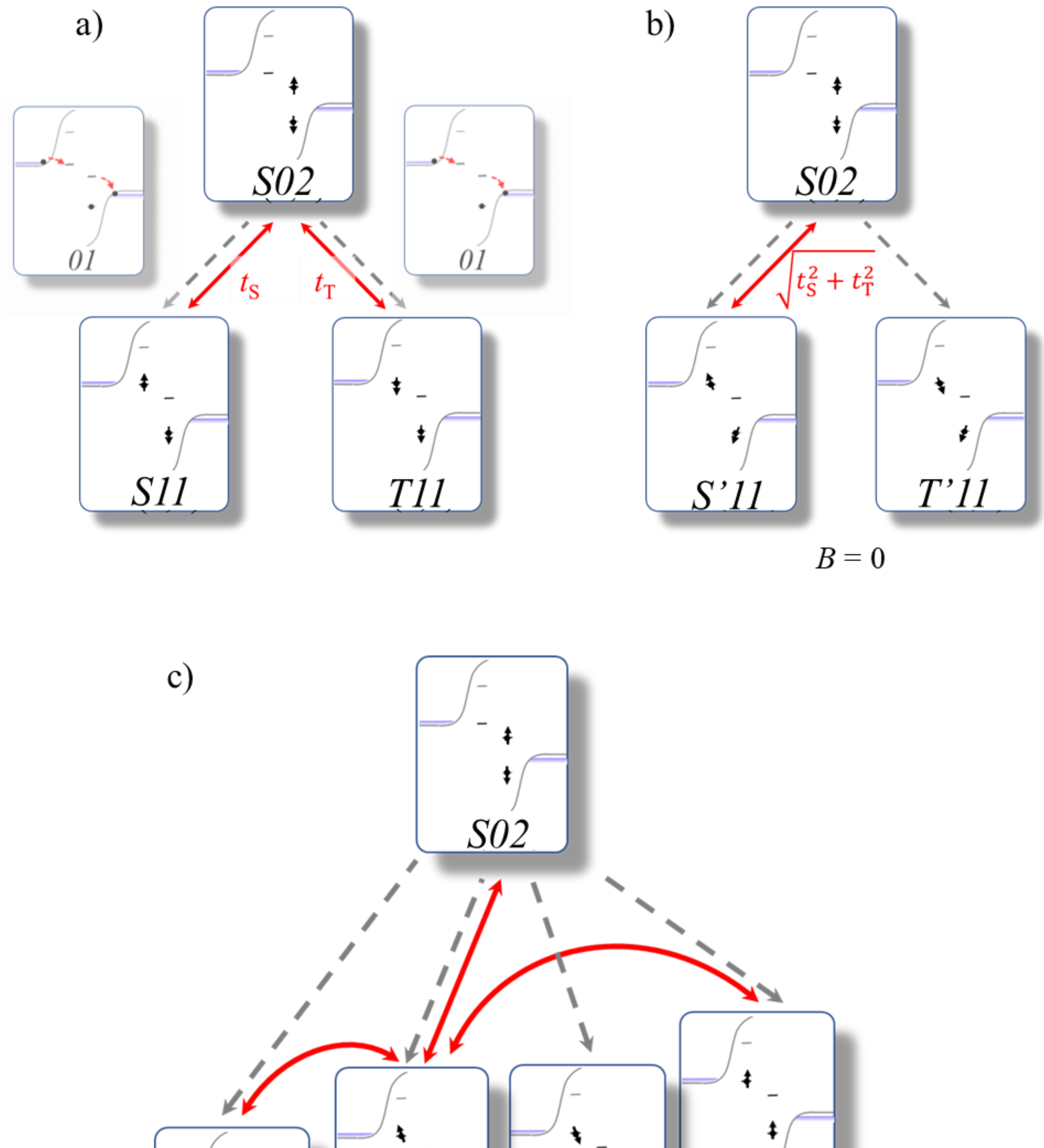

**Supplementary Figure 8.**

PSB as a lambda system. **a** Schematic of the three charge spin states *S02*, *S11*, and *T11* in PSB and the transitions between them. **b** *S11* and *T11* states are rotated to become *S'11* and *T'11*. For a suitable rotation angle, CPT at *T'11* will occur. **c** In the magnetic field, $T_{-}11$ and $T_{+}11$ undergo Zeeman shifts, and the CPT of these two states is lifted.

| Figure | Dopant | $L_{eff}$ (nm) | $W$ (μm) | $V_{SD}$ (V) | $V_G$ (V) | $I_{SD}$ (pA) | SR (%) | Dip (mT) | Peak (mT) | Remarks |
|---|---|---|---|---|---|---|---|---|---|---|
| 2a and 2b | S | 22 | 5 | | | | | | | SCT 10 and 300 K |
| 2d and 2e | Zn | 16 | 5 | | | | | | | SCT 10 and 300 K |
| S3a and S3b | S/Zn | 14 | 5 | | | | | | | SCT 10 and 300 K |
| 3e | S/Zn | 20 | 1 | 0.5 | -0.12 | 515.5 | 1.2 | 191 | | PSB at 300 K |
| 4a | S/Zn | 22 | 10 | 0.4 | 0.38 | 10.88 | 1.3 | 1.8 | 22 | PSB at 300 K |
| 4b | S/Zn | 16 | 10 | 0.3 | 0.42 | 5.272 | 0.4 | 0.26 | 15.5 | PSB at 300 K |
| S4a | S/Zn | 16 | 5 | 0.45 | -0.05 | 318.8 | 3.5 | 190 | | PSB at 300 K |
| S4b | S/Zn | 18 | 10 | 0.6 | -0.23 | 110.1 | 0.4 | 84 | | PSB at 300 K |
| S4c | S/Zn | 18 | 1 | 0.3 | 0.6 | 10.51 | 1.1 | 2.5 | 26.3 | PSB at 300 K |
| S4d | S/Zn | 18 | 5 | 0.45 | 0.35 | 7.095 | 1.4 | 2 | 21 | PSB at 300 K |
| S4e | S/Zn | 24 | 10 | 0.3 | 0.425 | 6.424 | 0.4 | 1.67 | 27.7 | PSB at 300 K |
| 5a-5c | S/Zn | 18 | 5 | 0.75 | 0.28 | 1.4 | | | | MR up to 50 K |
| 5d and 5e | S/Zn | 24 | 10 | 0.799 | 0.376 | 0.56 | | | | Rabi at 10 K |

**Supplementary Table 1.**

Parameters for the 13 devices discussed in the manuscript. Column description—Figure: figure corresponding to data from each device (S: Supplementary Figure); Dopant: distinguishes doping conditions among “only S,” “only Zn,” or “both of S and Zn”; $L_{eff}$: effective channel length; $W$: channel width; $V_{SD}$ and $V_G$: values of $V_{SD}$ and $V_G$ used for each figure; $I_{SD}$: $I_{SD}$ values at zero $B$; SR: signal ratio (the change in $I_{SD}$ within the measured $B$ range is normalized using $I_{SD}$); Dip: width of zero-B dip. Peak: width of the zero-$B$ peak envelope; Remarks: phenomena observed (single charge tunneling (SCT), Pauli spin blockade (PSB), magnetic resonance (MR), and Rabi oscillation of the pulsed magnetic resonance (Rabi)).